\documentclass[letter,twocolumn]{jpsj3_pdftex}
\usepackage{amsmath}
\usepackage{amssymb}
\usepackage{bm}
\usepackage{color}
\usepackage{graphicx}
\usepackage{cite}
\usepackage[normalem]{ulem}

\definecolor{darkgreen}{rgb}{0,0.5,0}

\definecolor{purple}{rgb}{0.5,0,0.5}

\definecolor{orange}{rgb}{0.7,0.3,0}

\allowdisplaybreaks

\def\fig{./}

\title{%
Gapless Spin-Liquid Phase in an Extended Spin $1/2$ Triangular
Heisenberg Model
}
\author{
\name{Ryui Kaneko}$^1$,
\name{Satoshi Morita}$^2$,
and
\name{Masatoshi Imada}$^3$
}
\inst{
$^1$Institut f\"{u}r Theoretische Physik, Goethe-Universit\"{a}t Frankfurt,
Max-von-Laue-Stra\ss{}e 1, 60438 Frankfurt am Main, Germany
\\
$^2$Institute for Solid State Physics, University of Tokyo,
Kashiwa, Chiba 277-8581, Japan
\\
$^3$Department of Applied Physics, University of Tokyo,
Bunkyo, Tokyo 113-8656, Japan
}

\abst{%
We numerically study the Heisenberg models on triangular lattices
by extending it from the simplest equilateral lattice with only the
nearest-neighbor exchange interaction. We show that, by including an
additional weak next-nearest-neighbor interaction,
a quantum spin-liquid phase is stabilized against the antiferromagnetic
order.
The spin gap
(triplet excitation gap) and spin correlation at long distances decay
algebraically with increasing system size at the critical point between
the antiferromagnetic phase and the spin-liquid phase.
This algebraic behavior continues in the
spin-liquid phase as well, indicating
the presence of an unconventional critical
(algebraic spin-liquid) phase characterized by the dynamical and
anomalous critical exponents $z+\eta\sim1$.  Unusually small triplet and
singlet excitation energies found in extended points of the Brillouin
zone impose constraints on this algebraic spin liquid.
}

\begin{document}

\maketitle


The spin liquid was first proposed by Anderson as a resonating valence
bond (RVB) ground state of the spin $1/2$ antiferromagnetic Heisenberg
model on the triangular lattice~\cite{0025-5408(73)90167-0}.  Although
it is now widely believed that the true ground state of the model has a
long-ranged magnetic order~\cite{PhysRevB.50.10048,
PhysRevLett.82.3899,PhysRevLett.99.127004}, two-dimensional systems
based on some modification of the triangular Heisenberg model provide us
with realistic possibilities of reaching the spin liquid.  This is
because some organic conductors are basically described by the
triangular structure with some additional elements, and they indeed do
not show clear indications of symmetry
breaking~\cite{PhysRevLett.91.107001, PhysRevB.77.104413}.  If one can
understand how and in which direction the spin-liquid state becomes more
stabilized and what kind of spin liquids is expected even at the level
of model studies, it will greatly help as a clue for stabilizing spin
liquids.

\begin{figure*}[htb]
\centering
\includegraphics[width=.9\linewidth]{\fig/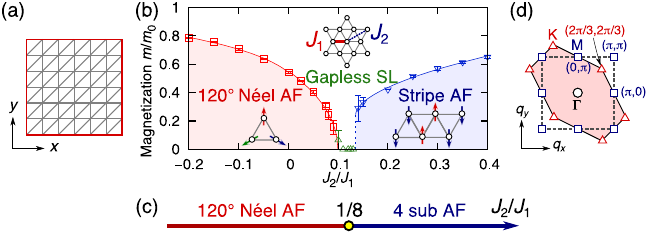}
\caption{(color online)
(a)
Deformed triangular lattice, which is topologically equivalent to
isotropic triangular lattice.  (b) Phase diagram of
antiferromagnetic $J_1$-$J_2$ Heisenberg model on triangular lattice
obtained in the present study.  The red squares, green up-pointing
triangular points, and blue down-pointing triangular points are order
parameters (sublattice magnetizations) of the 120$^\circ$ N\'eel states
(120$^\circ$ N\'eel AF), spin-liquid (SL) states, and stripe states (stripe AF),
respectively.  The lines are guides for the eyes.  The magnetization is
estimated as $m/m_0 = 2\sqrt{c\times
\lim_{N_{\mathrm{s}}\rightarrow\infty} S(\bm{Q}) /N_{\mathrm{s}} }$,
where $c$ denotes a correction factor, and $S(\bm{Q})$ is the peak value
of the spin structure factor.  For the 120$^\circ$ N\'eel and
SL states, $c=2$,
which is compatible with the standard
definition~\cite{PhysRevB.50.10048,
PhysRevLett.82.3899,PhysRevLett.99.127004},
while for the stripe states, $c=1$.  Here,
$m_0(=1/2)$ is the saturated magnetization expected in the classical
N\'eel order.  
(c) Phase diagram of classical model as reference.
The transition from the 120$^\circ$ N\'eel AF state to the four-sublattice
antiferromagnetic state (4 sub AF) occurs at $J_2/J_1=1/8$.
The stripe AF state is one of the 4 sub-AF states, and
it is selected under quantum fluctuations.
(d) Brillouin zone of deformed
triangular lattice.  Equivalent momenta are represented by the same
symbols.
}
\label{fig:j1j2_schematic_PD}
\end{figure*}

In this letter, we study the Heisenberg model on the triangular lattice
shown in Fig.~\ref{fig:j1j2_schematic_PD}(a) with
a weak next-nearest-neighbor interaction $J_2$ illustrated in the inset of
Fig.~\ref{fig:j1j2_schematic_PD}(b).  The Hamiltonian
has the form
\begin{equation}
 H = J_1 \sum_{\left<i,j\right>}
     \bm{S}_i\cdot\bm{S}_j
   + J_2 \sum_{\left<\!\left<i,j\right>\!\right>}
     \bm{S}_i\cdot\bm{S}_j,
\end{equation}
where $\bm{S}_i$, $\left<i,j\right>$, and
$\left<\!\left<i,j\right>\!\right>$ denote the quantum $S=1/2$ spin at
site $i$, nearest-neighbor sites, and next-nearest-neighbor sites,
respectively.

Previous studies~\cite{PhysRevLett.63.2524, PhysRevB.47.9105,
PhysRevB.47.6165, PhysRevB.60.9489, PhysRevB.42.4800, PhysRevB.46.11137,
BF01308811, PhysRevB.52.6647, 0953-8984-7-46-010,
PhysRevLett.111.157203} have suggested that the above model contains rich
phases, as shown in Fig.~\ref{fig:j1j2_schematic_PD}(c).
For the classical Heisenberg model, the 120$^\circ$ N\'eel
state illustrated in Fig.~\ref{fig:j1j2_schematic_PD}(b)
is stable for $J_2/J_1<1/8$, while for $J_2/J_1>1/8$ the ground
states become degenerate for any four-sublattice states that satisfy
$\bm{S}_{1} + \bm{S}_{2} + \bm{S}_{3} + \bm{S}_{4} = 0$, where
$\bm{S}_i$ denotes the spin at sublattice site $i$.  When $J_2/J_1>1$,
an incommensurate spin structure appears.  By considering quantum
fluctuations perturbatively, the stripe-type antiferromagnetic state
illustrated in
Fig.~\ref{fig:j1j2_schematic_PD}(b)~\cite{PhysRevB.47.9105,
PhysRevB.47.6165, PhysRevB.60.9489,
PhysRevB.42.4800, PhysRevB.46.11137, BF01308811, PhysRevB.52.6647} was
proposed for $1/8<J_2/J_1<1$.  Then, the spin-liquid state could emerge
at approximately $J_2/J_1=1/8$, sandwiched by the 120$^\circ$ N\'eel and stripe
states~\cite{PhysRevB.47.9105,PhysRevB.60.9489}.  Very recently, by
using a variational Monte Carlo (VMC) method, Mishmash {\it et al.}
claimed that the spin-liquid phase with the nodal $d$-wave symmetry is
realized for $0.05\lesssim J_2/J_1\lesssim
0.17$~\cite{PhysRevLett.111.157203}.  However, they overestimated the
spin-liquid phase, since their Huse-Elser-type wave functions for
antiferromagnetic states are relatively inaccurate.  Moreover, their wave
function favors the $d$-wave-type spin liquid due to a limited number of
variational parameters.

To reveal the nature of the spin-liquid ground states possibly realized
in this model, we calculate the ground states and low-energy excitations
up to $18\times 18$ sites
by a many-variable variational Monte Carlo (mVMC)
method~\cite{JPSJ.77.114701}
(see Supplemental Materials~\cite{supp1} for more details).
We find a spin-liquid phase characterized
by gapless excitations and the power-law decay of the spin correlation
with the critical and dynamical exponents $\eta$ and $z$,
respectively, satisfying $z+\eta = 1$ in the parameter region $0.10(1)\le
J_2/J_1\le 0.135(5)$, as shown in
Fig.~\ref{fig:j1j2_schematic_PD}(b).

\begin{figure*}[htb]
\centering
\includegraphics[width=.9\linewidth]{\fig/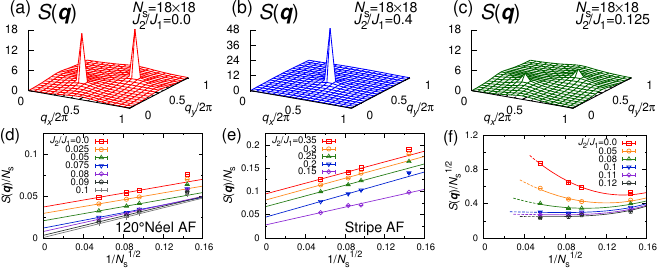}
\caption{(color online)
(a-c) Spin structure
factors ($N_{\mathrm{s}}=18\times 18$) for 120$^\circ$ N\'eel state at
$J_2/J_1=0$ (a), stripe state at $J_2/J_1=0.4$
(b), and spin-liquid state with short-ranged
magnetic correlation with the 120$^\circ$ spin alignment at
$J_2/J_1=0.125$ (c).  (d-f) Size dependences of peak value of
$S(\bm{Q})/N_{\mathrm{s}}$ for 120$^\circ$ N\'eel states (except
$J_2/J_1=0.1$) (d), stripe states
(e), and $S(\bm{Q})/N_{\mathrm{s}}^{1/2}$ for
120$^\circ$ N\'eel states and spin
liquid states (f).  Size dependence at $J_2/J_1=0.1$ is
also shown in (d) for reference.  We extrapolate the
data using the relation $S(\bm{Q})/N_{\mathrm{s}} = m^2/c +
S_1/N_{\mathrm{s}}^{1/2} +S_2/N_{\mathrm{s}} $, where $m$ denotes the
extrapolated magnetization, $c$ denotes a correction factor, which is the
same as in Fig.~\ref{fig:j1j2_schematic_PD}(a), while
$S_1$ and $S_2$ are constants.  For antiferromagnetic states, we use the
data for $N_{\mathrm{s}}>100$ to reduce the finite-size effect, and fit
them by the relation with $S_2=0$ to estimate the statistical error
bars.  On the other hand, to precisely determine the phase boundary for
spin-liquid states, we also examine the size dependence of $S(\bm{Q})$.
We find that $S(\bm{Q})$ scales as $S(\bm{Q})\propto
N_{\mathrm{s}}^{1/2}$ for $J_2/J_1\ge 0.1$.
}
\label{fig:j1j2_sqpeak_fit}
\end{figure*}

Three locally stable states are found as candidates of the
ground state: the 120$^\circ$ N\'eel, stripe, and
spin-liquid states.  As shown in Fig.~\ref{fig:j1j2_sqpeak_fit}, these
states are characterized by the spin structure factor defined as
\begin{equation}
 S(\bm{q}) = \frac{1}{ N_{\mathrm{s}} } \sum_{i,j}
 e^{i\bm{q}\cdot(\bm{R}_i-\bm{R}_j)}
 \left< \bm{S}_i\cdot\bm{S}_j \right>,
 \label{eq:j1j2_S(q)}
\end{equation}
where $\bm{R}_{i}$ denotes the position vector of site $i$.  The
existence of the long-ranged order is ensured when the maximum of
$S(\bm{q})$ at $\bm{q}=\bm{Q}$ scales as $S(\bm{Q})\propto
N_{\mathrm{s}}$ for a large $N_{\mathrm{s}}$.  The ground state is
determined from the comparison of the energy, which is extrapolated to
the thermodynamic limit, if more than one locally stable states are
found.

For $J_2/J_1<0.1$, we find the 120$^\circ$ N\'eel state as the ground
state. In the Brillouin zone illustrated in
Fig.~\ref{fig:j1j2_schematic_PD}(d), Bragg peaks
appear at $\bm{q}=\pm(2\pi/3,2\pi/3)$ corresponding to the 120$^\circ$
spin structure, as shown in
Fig.~\ref{fig:j1j2_sqpeak_fit}(a).  From the size
extrapolation of $S(\bm{Q})/N_{\mathrm{s}}$ in
Fig.~\ref{fig:j1j2_sqpeak_fit}(d), we obtain the
magnetization of the 120$^\circ$ N\'eel state at $J_2/J_1=0$ as $m/m_0 =
0.543(6)$, where $m_0=1/2$ is the saturated magnetization in spin $1/2$
systems.  This value is nearly the same as the
best variational result ($m/m_0=0.53$)
in the literature~\cite{PhysRevB.80.012404}.
For $J_2/J_1>0.135$, we find the stripe
ordered state as the ground state.  As shown in
Fig.~\ref{fig:j1j2_sqpeak_fit}(b), a Bragg
peak appears at $\bm{q}=(\pi,\pi)$,
corresponding to the ferromagnetic spin
alignment along the $x=y$ direction and the antiferromagnetic spin
alignment along the $x$- and $y$-axes on the lattice illustrated in
Fig.~\ref{fig:j1j2_schematic_PD}(a).  The state has
a long-ranged order, as shown in
Fig.~\ref{fig:j1j2_sqpeak_fit}(e).

Sandwiched by the 120$^\circ$ N\'eel and stripe states, the spin liquid
is stabilized.  
Diffusive peaks appear at $\bm{q}=\pm(2\pi/3,2\pi/3)$,
reminiscent of the 120$^\circ$ spin alignment, as shown in
Fig.~\ref{fig:j1j2_sqpeak_fit}(c).  We find in
Fig.~\ref{fig:j1j2_sqpeak_fit}(f) that the peak
scales as $S(\bm{Q})\propto N_{\mathrm{s}}^{1/2}$ in all the regions of
the spin-liquid phase ($0.10<J_2/J_1<0.135$), implying the emergence of a
critical phase without the long-ranged order.  In fact, this scaling
leads to the spin correlation $\langle \bm{S}_i\cdot
\bm{S}_j\rangle\propto 1/|\bm{R}_i-\bm{R}_j|$~{\cite{supp2}}.

At the transition point $J_2/J_1=0.10(1)$, the spin liquid undergoes a
continuous quantum phase transition to the 120$^\circ$ N\'eel state,
while at $J_2/J_1=0.135(5)$ a level crossing
to the stripe state occurs.

\begin{figure*}[htb]
\centering
\includegraphics[width=.9\linewidth]{\fig/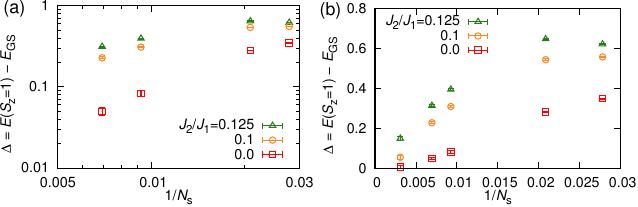}
\caption{(color online)
Size scalings of spin gaps of 120$^\circ$ N\'eel
state, spin liquid, and near the critical point.
The spin gaps at the
$\Gamma$ point are plotted on a log-log plot (a)
and a linear plot (b).
}
\label{fig:j1j2_spingap_fit}
\end{figure*}

\begin{figure*}[htb]
\centering
\includegraphics[width=.9\linewidth]{\fig/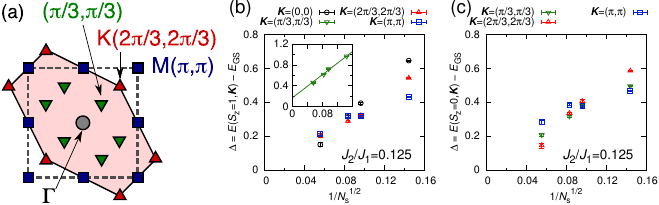}
\caption{(color online)
Momentum-resolved excitation spectra in spin-liquid
phase.
(a) Momenta studied for excitation spectra
in colors the same as in (b) and (c).
Size dependences of triplet (b) and singlet
(c) gaps of each total momentum for spin-liquid
phase ($J_2/J_1=0.125$).  Triplet excitations are gapless only at the
$\Gamma$ and K points for the 120$^\circ$ N\'eel state, while they are
gapless only at the $\Gamma$ and M points for the stripe state.  For the
spin liquid, in addition to the K, M, and $\Gamma$ points, the middle
points of the
$\Gamma$ and K points show unusually small excitation energies.
}
\label{fig:j1j2_BZ_triplet_min} 
\end{figure*}

To get further insight into the nature of the spin-liquid ground state,
we study the energies of the momentum-resolved lowest triplet and
singlet excited states.  We take the total momenta along the symmetric
$\Gamma$-K-M line, namely, $\bm{K}=(q,q)$ with $0\le q\le \pi$, as
illustrated in Fig.~\ref{fig:j1j2_schematic_PD}(d).
The ground states have turned out to be always the singlet and stay at
the $\Gamma$ point irrespective of $J_2$ and $N_{\mathrm{s}}$.

According to the results of spin wave (SW) analyses~\cite{BF01309171,
PhysRevB.39.2608}, the spin gap $\Delta$ for antiferromagnetic states
should scale as $\Delta(N_{\mathrm{s}})\propto 1/N_{\mathrm{s}}$.
However, when both the magnetic order and spin gap become zero, especially
on continuous transition points, a more subtle scaling dominates.  For
example, for the two-dimensional quantum Heisenberg model, when a
quantum phase transition occurs between the phase with an $\mathrm{O}(3)$
long-ranged order and the disordered phase with a finite-gap spin
excitation, the nature of the transition was proposed to be essentially
the same as that of the two-dimensional quantum $\mathrm{O}(3)$
nonlinear sigma model~\cite{JPSJ.66.2957}.  Since the correlation length
is the same in the spatial and time directions, the dynamical exponent
$z$ becomes $z=1$ at the critical point.  Therefore, the spin gap, which
scales as $\Delta\sim k^z$ at the critical point~\cite{1.3518900}, obeys
the relation $\Delta\sim k\propto 1/L$ with $L\equiv
N_{\mathrm{s}}^{1/2}$.  Later on, a possible deconfined criticality was
proposed in that case~\cite{science.1091806,PhysRevB.70.144407}.

In general, the critical state may be characterized by the power-law
decay $\Delta\propto 1/L^{z}$ with the dynamical exponent $z$.  As shown
in Fig.~\ref{fig:j1j2_spingap_fit}, the spin gap at the critical point
($J_2/J_1=0.1$) consistently shows power-law decay for larger sizes,
although it is difficult to estimate the critical exponent of the decay
$z$ quantitatively.  In finite-size systems, higher-order corrections in
the extrapolation of the spin gap are nonnegligible, which makes the
extrapolation difficult.  Indeed, in a previous study
on the triangular lattice system, the spin gap of
the spin liquid 
scales
as a concave function of $1/N_{\mathrm{s}}$ if the system size is not
sufficiently large~\cite{PhysRevB.74.014421}, although the chiral
perturbation theory for the square lattice with the well-established
antiferromagnetic order predicts the scaling by $1/N_{\mathrm{s}}$ with
the negative coefficient of the $1/N_{\mathrm{s}}^{3/2}$
correction~\cite{BF01309171}. Therefore, the estimated critical
exponents may become larger than the exact exponents.

This critical point turns out to be unusual, because the spin gap
appears to decay as a power law even deep inside the spin-liquid phase.
Within the available system size ($\le 18\times 18$), these results
support the nation that the whole spin-liquid phase may belong to the critical
phase.  This is consistent with the above critical behavior of the spin
correlation, because we expect $\langle \bm{S}_i\cdot
\bm{S}_j\rangle\propto 1/|\bm{R}_i-\bm{R}_j|^{d+z+\eta-2}$ with $d=2$
and $z+\eta=1$ in the whole spin-liquid phase,
as we will detail later.

We now study the momentum-resolved spectra of the excitations
$\Delta(\bm{K})$.  Generally, if the ground state has a magnetic
long-ranged order, corresponding to the magnon modes, it shows gapless
excitations at the $\Gamma$ point as well as at the wave vector
$\bm{Q}$ of $S(\bm{Q})$ peaks.  We have indeed confirmed that the
present mVMC calculation reproduces this expectation.

In the spin-liquid phase, we examine the size dependence of singlet and
triplet gaps at $\bm{K}=(n\pi/3,n\pi/3)$ ($n\in \mathbb{Z}$), as shown in
Fig.~\ref{fig:j1j2_BZ_triplet_min}(a), which can be
calculated for all the system sizes that we studied~\cite{supp3}
(see also Supplementary Materials for data at other $k$ points).
As shown in
Figs.~\ref{fig:j1j2_BZ_triplet_min}(b) and
\ref{fig:j1j2_BZ_triplet_min}(c),
we find that both singlet and triplet gaps at
all the available momenta systematically and substantially decrease
with increasing system size, and natural extrapolations suggest
vanishingly and unusually small values if it is nonzero. Even at
$\bm{K}=(\pi/3,\pi/3)$, where the gap is largest, a simple linear
extrapolation suggests the gap $<0.2J_1$. The gaps at other momenta are
likely to be less than $0.1J_1$. To our knowledge, such small gaps
in the extended points of the Brillouin zone are not known. In the
antiferromagnetic phase, the spin wave excitation usually has
dispersions of the order of the exchange coupling $J_1$.

Now, we discuss comparisons with previously proposed spin-liquid states.
In the present result, the singlet and triplet excitation energies look
vanishingly small at all the studied total momenta, including the
$\Gamma$, K, and M points, as well as the middle of the $\Gamma$ and
K points, as shown in
Fig.~\ref{fig:j1j2_BZ_triplet_min}(a).  If the
excitation is gapless at the momenta in a finite area of the Brillouin
zone, it can be accounted for by the presence of a large spinon Fermi
surface~\cite{PhysRevB.72.045105, PhysRevLett.95.036403}.  Although the
present model does not contain itinerant electrons, if the finite-size
spin gap scales as $1/L$, the present spin-liquid state is indeed
similar to the expected gap scaling of the lowest particle-hole (Stoner)
excitation in the (spinon) Fermi liquid.  However, the exponent of the
power-law decay of the spin correlation ($\sim 1/r^{\alpha}$ with
$\alpha=1$) contradicts the expected behavior in the presence of the
Fermi surface, namely, $\left<\bm{S}_i\cdot \bm{S}_j\right> \propto
1/|\bm{R}_i-\bm{R}_j|^3$~\cite{JPSJ.61.3331}.

When the gap closes only at $\bm{K}=(n\pi/3,n\pi/3)$ ($n\in \mathbb{Z}$)
or at one of these $\bm{K}$ points,
it is consistent with the algebraic spin liquid.  For
example, for the Dirac spin liquid whose dispersion scales as
$\Delta\propto k^z$ with $z=1$, it was discussed that the spin
correlation may decay as $1/r^{1+\eta}$ typically with a nonzero $\eta$
because of the coupling of Dirac spinons to a noncompact U(1) gauge field
(see Refs.~\citenum{PhysRevLett.86.3871} and \citenum{PhysRevB.72.104404}
and references therein).
The coupling depends on the flux pattern of the slave-boson
mean-field states, and thus the exponent $\eta$ varies according to
lattice model; for instance, it was proposed that $\eta\sim 0.5$ for
the projected $d$-wave BCS state~\cite{PhysRevB.72.104404} and $\eta\sim
3$ for the $\mathrm{U(1)}$ Dirac spin liquid on the kagome
lattice~\cite{PhysRevB.77.224413}.  More generally, it can be
conjectured that the algebraic spin liquid is characterized by the
$1/L^z$ scaling of the gap closing and the $1/r^{z+\eta}$ decay of the
spin correlation.  If $z+\eta=1$, it is consistent with the present
result.  The vanishingly small excitation energies at extended momentum
points naturally lead to the power-law decay of the spin correlation in
real space at the period of the corresponding momentum points.  However,
so far, the clear algebraic ($1/r$) decay of the spin correlation is
found only for that at the period $(2\pi /3,2\pi /3)$ (120$^\circ$ order
point).  This rules out the possibility of a $d$-wave-type spin
liquid~\cite{PhysRevB.81.245121, PhysRevLett.111.157203}, whose gap is
open at the period $(2\pi /3,2\pi /3)$.  In the present calculation, the
spin correlations of the period at other momentum points decay at least
faster than $1/r^2$ because $S({\bm q})$ is at most logarithmically
divergent and is likely to stay finite in the thermodynamic limit.  If
this picture applies so that some of the spin correlations at the
periods other than $(2\pi /3,2\pi /3)$ decay exponentially, we expect
the opening of gaps at the wavenumber away from the crossing points.  So
far, the examination of this prediction is not numerically easy with the
present computer power. This is left for future studies. Within the
available results, the present spin liquid does not contradict the
possible algebraic spin liquid or flux state, while our results impose
constraints on its nature.

We have shown that the spin-$1/2$ Heisenberg model on the triangular
lattice with the nearest and next-nearest exchange interactions shows an
unusual spin liquid, where the spin correlation decays as $1/r$ and the
finite-size gap closes as a power law as if the quantum critical point is
extended to a critical phase. Unusually small gaps at all the total
momenta studied here, however, impose a severe constraint on the nature
of the spin liquid found in the present study.  Clearly, further studies
are needed for a more unified understanding of the spin-liquid states
found in the present study.


\section*{Acknowledgements}

The mVMC codes used for the present computation were based on those first
developed by Daisuke Tahara and M. I. and those extended by S. M.
and Moyuru Kurita.
To compute the Pfaffian of skew-symmetric matrices, we employed
PFAPACK~\cite{2331130.2331138}.
The authors thank Takahiro Misawa for fruitful discussions.
This work was financially supported by MEXT HPCI Strategic Programs for
Innovative Research (SPIRE) and Computational Materials Science
Initiative (CMSI). The numerical calculation was partly carried out at the
Supercomputer Center, Institute for Solid State Physics, University of Tokyo.
The numerical calculation was also partly carried out at K computer at RIKEN
Advanced Institute for Computational Science (AICS) under grant numbers
hp120043, hp120283, and hp130007.  This work was also supported by a
Grant-in-Aid for Scientific Research (Nos. 22104010 and 22340090)
from MEXT, Japan.




\end{document}



\vspace{\baselineskip}%
\noindent {\bfseries
Supplemental Materials to ``Gapless spin-liquid phase in an extended
spin $1/2$ triangular Heisenberg model''}

\section*{Variational Monte Carlo method}

We consider the variational wave function in the form $\left|\Psi\right>
= \mathcal{P} \mathcal{L} \left| \Psi_0\right> $ with
\begin{equation}
 \left| \Psi_0\right>\equiv
 \left(
 \sum_{i,j=1}^{N_{\mathrm{s}}}
 \sum_{s,t=\uparrow~\mathrm{or}~\downarrow} f_{ij}^{st}
 c_{is}^{\dagger} c_{jt}^{\dagger}
 \right)^{N_{\mathrm{e}}/2}
 \left|0\right>,
\end{equation}
which can describe antiferromagnetic states with collinear,
noncollinear, and even noncoplanar spin alignment, and also gapless and
gapped spin-liquid states on an equal footing.  Here, $\left|
\Psi_0\right>$ is a generalized Bardeen-Cooper-Schrieffer (BCS) wave
function, where $f_{ij}^{st}$ and $\left|0\right>$ denote the
variational parameter and the vacuum.  This is a natural extension of
the Liang-Doucot-Anderson type wave function~\cite{PhysRevLett.61.365}
and the Hartree-Fock-Bogoliubov type wave
function~\cite{PhysRevB.43.12943, PhysRevLett.85.4345}.  Note that not
only the product of the paired singlet but also the paired triplet
component are allowed in this scheme beyond the RVB
singlet~\cite{PhysRevLett.61.365}, which is crucial in representing
noncollinear spin order.  We assume $f_{ij}^{st}$ to be long-ranged and
allow symmetry breaking, namely six-sublattice structure.  To restore
the translational symmetry of $\left|\Psi_0\right>$, we consider the
total momentum projection
\begin{equation}
 \mathcal{L}^{\bm{K}}
 = \frac{1}{N_{\mathrm{s}}}\sum_{\bm{R}}
 e^{i\bm{K}\cdot\bm{R}} T_{\bm{R}},
\end{equation}
where $\bm{K}$ and $T_{\bm{R}}$ denote the total momentum [defined in
Fig.~1(d)] and the
translation operator, respectively.  We apply the projection onto all
the possible total momenta.  In addition to the Gutzwiller factor
\begin{equation}
 \mathcal{P}_{\mathrm{G}} = \prod_{i}
 \left(1-n_{i\uparrow}n_{i\downarrow}\right)
\end{equation}
with $n_{is}=c_{is}^{\dagger}c_{is}$, which excludes the double
occupation of the electron, to reach a wave function closer to the true
ground state (lowest triplet state), we also impose a projection
$\mathcal{P}_{S^{z}_{\mathrm{tot}}=0}$
($\mathcal{P}_{S^{z}_{\mathrm{tot}}=1}$) on the wave function by keeping
the difference in the number of up and down spins fixed as zero (two).
We consider lattices for $N_{\mathrm{s}}=L\times L$ and $\sqrt{3}L\times
\sqrt{3}L$ 
up to $18\times 18$ sites
under the periodic boundary condition.  By using the
stochastic reconfiguration (SR) method~\cite{PhysRevLett.80.4558}, we
optimize the parameters by typically 2000 SR steps.  After confirming
the energy convergence, we calculate the expectation values of the
physical quantities and average them over more than 10 independent runs
containing typically 2000 Monte Carlo steps each to estimate statistical
errors.

\begin{figure}[htb]
\centering
\includegraphics[width=.9\linewidth]{\fig/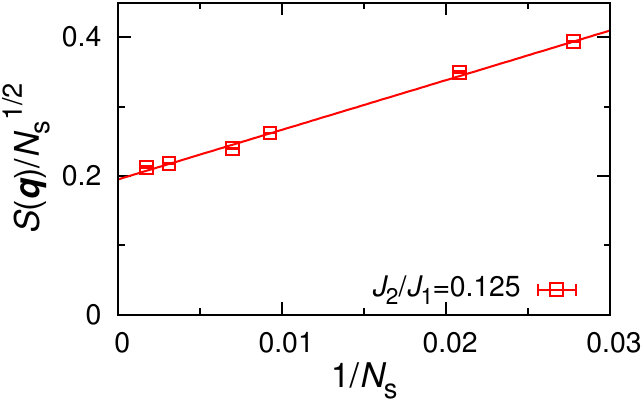}
\caption{(color online)
Size dependence of peak value of spin structure
factor divided by square root of system size. 
The data are calculated up to $24\times 24$ sites at
$J_2/J_1=0.125$.
It is well fit by the
scaling $S(\bm{q})/N_{\mathrm{s}}^{1/2} = S_1 + S_2/N_{\mathrm{s}}$,
where $S_1$ and $S_2$ are constants.
}
\label{fig:j1j2_sqpeakdivL_vs_N}
\end{figure}

\section*{Spin correlation function in spin-liquid phase}   

Here, we discuss the spin correlation function in the spin-liquid phase.
In the Ginzburg-Landau theory, the spin correlation function at the
critical point is given as $C(r)\propto r^{-(d-2)}$, where $d(>2)$ is
the spatial dimension.  Beyond the Ginzburg-Landau theory, at the
critical point and inside the critical phase, the correction may be
given by the dynamical exponent $z$ and the anomalous dimension $\eta$
as $C(r)\propto r^{-(d+z+\eta-2)}$
$(d+z+\eta>2)$~\cite{book.Goldenfeld,book.Cardy}.  This also leads to
$S(Q,N_{\mathrm{s}})\propto N_{\mathrm{s}}^{2-(d+z+\eta)/2}$ for
$d+z+\eta<4$.  Figure~S.\ref{fig:j1j2_sqpeakdivL_vs_N} shows that the spin
liquid follows the critical scaling with $z+\eta\sim 1$.

\begin{figure*}[htb]
\centering
\includegraphics[width=.9\linewidth]{\fig/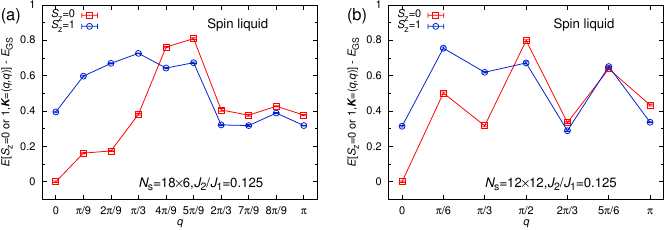}
\caption{(color online)
Total momentum dependence of singlet and triplet energies of spin
liquid for system sizes (a) $N_{\mathrm{s}}=6\sqrt{3}\times 6\sqrt{3}$
and (b) $N_{\mathrm{s}}=12\times 12$. 
The red squares and the blue circles
denote the singlet and triplet energies with the total momentum
$\bm{K}=(q,q)$. The lines are guide for the eye.
}
\label{fig:j1j2_other_gap}
\end{figure*}

\section*{Excitation energies at $k$-points other than
$\bm{k}=(n\pi/3,n\pi/3)$}

To understand the possible form of the excitation energy, we also
investigate the excitation energies for $N_{\mathrm{s}}=6\sqrt{3}\times
6\sqrt{3}$ and $12\times 12$ at the $\bm{k}=(q,q)$ with $q$ being
achievable wave vectors. As shown in Fig.~S.\ref{fig:j1j2_other_gap}, along the
$\bm{k}=(q,q)$ line for the finite-size systems, the singlet excitations
appear to have a peak around $\bm{k}=(\pi/2,\pi/2)$. On the other hand,
the triplet excitations appear to have a broad beak around
$\bm{k}=(\pi/3,\pi/3)$. The size of these excitation energies are less
than or nearly comparable to the larger triplet gap at
$\bm{k}=(\pi/3,\pi/3)$. Although it is difficult to estimate the
dispersion in the thermodynamic limit, since even the larger triplet gap
at $\bm{k}=(\pi/3,\pi/3)$ converges to a tiny value ($\Delta<0.2J_1$),
it is likely that almost all of the  excitations along $\bm{k}=(q,q)$
line converges to a near zero value. In this case, it can be conjectured
that nearly dispersion-less excitations appear along the $\Gamma$-K
lines and the K-M lines.


